\documentclass[aps,prl,twocolumn,groupedaddress]{revtex4}

\usepackage{graphicx}

\begin{document}

\title{Low-Noise Amplification of a Continuous Variable Quantum State}
\author{R. C. Pooser$^1$}\email{rpooser@nist.gov}
\author{A. M. Marino$^1$}
\author{V. Boyer$^{1,2}$}
\author{K. M. Jones$^3$}
\author{P. D. Lett$^1$}\email{paul.lett@nist.gov}
\affiliation{$^1$Joint Quantum Institute, National Institute of
Standards and Technology and the University of Maryland,
Gaithersburg, MD 20899 USA\\
$^2$MUARC, School of Physics and Astronomy, University of Birmingham, Edgbaston, Birmingham B15 2TT, UK\\
$^3$Department of Physics, Williams College, Williamstown, MA 01267 USA}

\date{\today}
\begin{abstract}
We present an experimental realization of a low-noise,
phase-insensitive optical amplifier using a four-wave mixing
interaction in hot Rb vapor. Performance near the quantum limit for
a range of amplifier gains, including near unity, can be achieved.
Such low-noise amplifiers are essential for so-called quantum
cloning machines and are useful in quantum information protocols. We
demonstrate that amplification and ``cloning'' of one half of a
two-mode squeezed state is possible while preserving entanglement.
\end{abstract}
\maketitle

The theory of an ideal, linear, phase-insensitive amplifier for an optical state is well developed \cite{Caves_amplifierPRD}. Such devices are of interest for implementing continuous variable (CV) quantum computing and quantum information protocols \cite{braunstien_rmp,Grangier_QKDarxiv,Messikh_JPhysB}, in particular as part of a quantum cloner designed to make the best possible copy of a quantum state \cite{Cerf_cloningPRL_2000,Braunstein_cloningPRL,Fiurasek_cloningPRL}. 
In this context a linear, phase-insensitive amplifier may be considered ``universal'' as its operation is independent of the quantum state of the input light.

Quantum mechanics predicts that any optical amplifier must add a
certain level of noise \cite{Caves_amplifierPRD} which insures that
such a device cannot be used to precisely clone an arbitrary quantum
state \cite{nocloning1,nocloning2,nocloning3,nocloning4}.  Amplifier
performance is often described in terms of the noise figure (NF),
which is the signal to noise ratio (SNR) of the amplified signal
divided by the input SNR:
NF$=\mathrm{SNR_{out}}/\mathrm{SNR_{in}}$. An ideal
quantum-noise-limited phase-insensitive amplifier, with a coherent state as the input, will have
$\mathrm{NF}=G/(2G-1)$,
where G is the intensity gain. Using such an amplifier and a beam
splitter one can produce multiple copies of the input which are called ``optimal quantum clones'' for arbitrary
Gaussian states. These are the best possible approximate copies of the original state \cite{Cerf_cloningPRL_2000, Hillary_cloningPRA_1996}.

While the theory of ideal quantum-noise-limited optical amplifiers
is well understood, practical implementations are few. Parametric
down conversion (PDC) in nonlinear crystals has been used to make
low-noise amplifiers, and Levenson, \textit{et al.} achieved near
quantum-noise-limited behavior in the high intensity pulsed pump
regime \cite{Levenson_amplificationJOSAB_1993}. In the CW pump
regime it was observed that PDC was quantum limited when coupling
efficiencies into the medium were accounted for
\cite{Ou_cloningPRL1993}. A completely different approach uses
linear optics and electronic feed forward techniques in order to
amplify \cite{Josse_amplificationPRL_2006} and optimally clone
\cite{Leuchs_cloningPRL2005} coherent states. Our experiment uses
near-resonant nondegenerate four-wave mixing (4WM) in $^{85}$Rb
vapor to amplify signals in a narrow-frequency band.
Although 4WM is often accompanied by sources of excess noise,
we have found conditions which allow the construction of a nearly
ideal, quantum-noise-limited amplifier.
By exploiting the low-noise characteristics of our device, we are able to amplify one of the modes from a
two-mode squeezed state (twin beams) in order to make
quantum clones. This represents an important step towards quantum cloning of an entangled state.

\begin{figure}[h]
\includegraphics[width=3in]{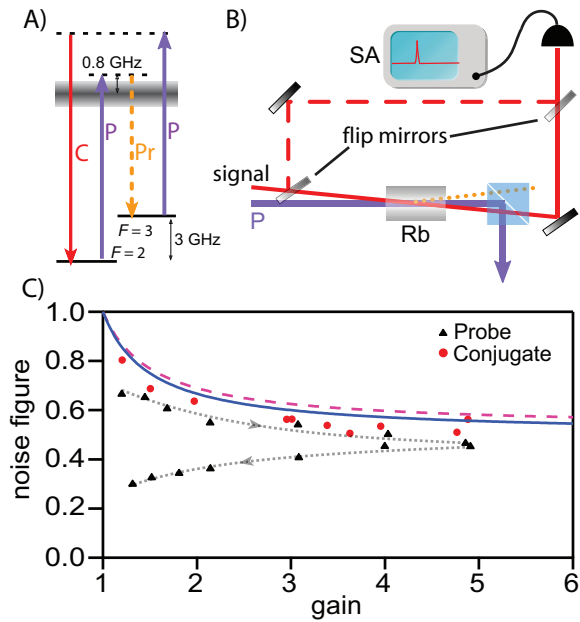}
\caption{(color online) A) Energy level diagram for the 4WM process. Pr: probe; C: conjugate; P: pump.  B) The configuration used to verify the noise figure of the amplifier. SA: spectrum analyzer. The dotted (orange) line represents the unused ancillary beam generated by the amplifier. C) The noise figure of the
amplifier for various gains. The solid (blue) curve represents the
ideal noise figure. The dashed (purple) curve shows the noise figure
that would be measured if an ideal amplifier were monitored with a
95\% efficient detector. The dotted line shows the change in gain as the pump frequency is moved from blue to red through the gain maximum for amplification of the probe beam.} \label{fig:NF}
\end{figure}

As a first step in characterizing the behavior of the 4WM-based
amplifier we  measure the NF as a function of gain when the input is a coherent state and compare this to the quantum-noise-limited case. Following the method of
Ref.~\cite{Leuchs_amplificationJMOD_2007} the input state is a 50 $\mu$W shot noise limited beam amplitude-modulated at 1 MHz to provide a signal about 20 dB above the noise. The test
configuration consists of a $^{85}$Rb vapor cell with a strong pump
injected along with the modulated input signal at a slight angle. There are
two input ports, either of which can be seeded while the other is
left with vacuum input. Depending on which port is seeded, the
signal is either upshifted or downshifted $\approx$3 GHz from the
pump, which is tuned near the D1 line (Fig.~\ref{fig:NF}A). Due to
4WM, the input is amplified, while a second output is produced at
the unseeded frequency \cite{squeezing_OL,Colin_squeezingPRA_2008}.
When only one input port is seeded the
process is phase-insensitive \cite{Science_images}, as will be shown
later. We call the frequency-upshifted beam the conjugate, and the
frequency-downshifted beam the probe.

The NF was calculated by comparing the SNR, which was measured with a radio
frequency spectrum analyzer centered at 1 MHz, before and after the cell
for various gains. The input SNR was measured by bypassing the vapor
cell using flip mirrors (Fig.~\ref{fig:NF}B), so that it would not
be underestimated due to losses on the vapor cell windows. Figure \ref{fig:NF}C
shows the NF as a function of gain.

The performance of the amplifier can be changed using different
methods. The temperature of the vapor cell controls the Rb number
density, while the detuning of the pump from the D1 line along with
the frequency difference between the pump and signal change the
strength of the nonlinearity. Only some combinations of temperature
and detunings result in a near-ideal NF. In Fig.~\ref{fig:NF}C the
gain was controlled by changing the pump detuning while maintaining
the difference between pump and signal frequencies at 3036 MHz. The
gain increases as the pump frequency comes closer to the atomic
resonance. The NF when the probe frequency is seeded (triangles) follows slightly below
the ideal curve as the pump is tuned from the blue toward the
maximum of the 4WM gain. As the pump is tuned red of the gain
maximum the NF degrades because of absorption of the probe at these
frequencies. The total scanning range was
$\approx$1 GHz centered at the 4WM gain maximum. The circles
represent the NF when the conjugate frequency is seeded and the pump is tuned from the
blue to near the gain maximum. The conjugate NF is in general better
than the probe since it experiences less absorption over a wider
frequency range. The cell temperature was 110 $^{\circ}$C for all measurements.

The detector efficiency was $\approx$95\%. Imperfect detectors cause an
overestimation of the NF since they underestimate the noise added by
the amplifier \cite{Leuchs_amplificationJMOD_2007}. The dashed curve
in Fig.~\ref{fig:NF}C shows the ideal NF rescaled in order to account
for detector efficiency, $\eta$: $\mathrm{NF}= G / (2\eta G - 2\eta
+1)$. Note that the data represent the ``as built'' behavior of the
actual device without any corrections for imperfections. In particular, we do not correct for
losses on the cell windows ($\approx$2\% per window) or polarizer
($\approx$1\%).

This study of coherent state amplification establishes that the 4WM
amplifier represents a practical approximation to an ideal
quantum-limited amplifier for Gaussian CV states. We now explore the action
of this amplifier on non-classical states; in particular we use it
to amplify one mode of a two-mode squeezed state. 
We study the cloning operation on one half of an entangled state by using a configuration in which the amplifier is followed by a variable attenuator, whose output simulates one of the outputs of a beam splitter. By setting the gain-attenuation product to one we can study a range of cloning configurations, including asymetric clones (when clones have unequal intensities) and the usual case with an amplifier gain of 2 followed by a 50\% beamsplitter (when there are two clones with equal intensity). 
The last case is an implementation of the ``local e-cloner'' discussed
theoretically in Ref.~\cite{weedbrook_cloningPRA_2008}.

\begin{figure}[h]
\includegraphics[width=3in]{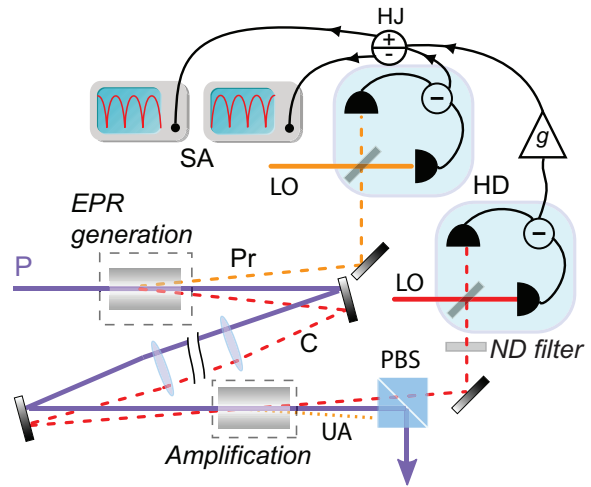}
\caption{(color online) Experimental setup: C: conjugate beam; Pr:
probe beam; SA: spectrum analyzer; PBS: polarizing beam splitter;
$g$: electronic attenuator; HJ: hybrid junction; LO: local
oscillator; HD: homodyne detector; UA: unused ancilla.
The LOs follow almost identical beampaths to those of the EPR beams
(dashed lines).} \label{fig:setup}
\end{figure}

\begin{figure}[h]
\includegraphics[width=3in]{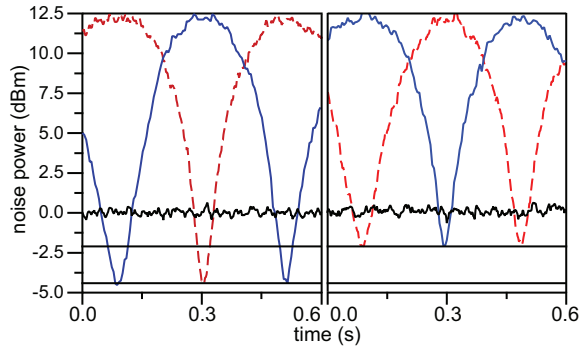}
\caption{(color online) Squeezing traces at 1 MHz (zero span, RBW: 10kHz, VBW: 300Hz) for the amplitude
difference and phase sum quadratures, normalized to the shot noise
level, for two different amplifier gains as a function of HD phase.
The HD phases are scanned synchronously in time so that they always
measure the same quadratures for each beam at a given time \cite{Science_images}. The traces on the left show the
squeezing level with the amplifier turned off and no ND filter in
the conjugate beam path. The right traces show the squeezing when
the amplifier gain is $\approx$1.8. The minima of the solid (blue) traces represent $\Delta \hat X_-^2$, while the minima of the dashed (red) traces represent $\Delta \hat Y_+^2$.}
\label{fig:squeezing}
\end{figure}

First, an initial vacuum two-mode squeezed state with 4.3(2) dB (all uncertainties are combined statistical and systematic) of noise reduction (see Fig.~\ref{fig:squeezing}) is
generated using 4WM starting with vacuum input in both ports
\cite{Science_images}.
As shown in Fig.~\ref{fig:setup}, after the
first cell the two output modes are separated. The conjugate is
passed along with the pump beam through a 4f imaging system and
input into a second $^{85}$Rb cell which acts as the amplifier. The
amplifier gain is controlled by adjusting the temperature of the
second cell, since the detunings of the various beams are
necessarily the same in both cells for our experimental setup.

Characterization of the entanglement relies on measurements of the
amplitude, $\hat X$, and phase, $\hat Y$, quadratures of each beam.
The quadratures are defined such that their variances for coherent
states (called the standard quantum limit, or SQL) are 1. Homodyne
detection, which requires mode matching the signal to a bright local
oscillator (LO), is
used to make the required measurements. To generate the LOs an identical 4WM process is used. This process has its own pump and is spatially separated from the twin beam generation but takes place in the same cell. It is seeded with a small input ($\approx 200 \mu$W), resulting in bright beams. The conjugate's LO (along with its own pump) is passed along with the conjugate and its pump through the 4f imaging system, is amplified in the second
cell, and is attenuated afterwards by a neutral density (ND) filter. The pumps for both LO and twin beams have equal powers. Generating LOs in this
way ensures that they have the same frequencies and wavefronts as
the twin beams \cite{Science_images}. Passing through the ND filter keeps the overall gain-loss product of the conjugate LO at unity,
thus the conjugate's homodyne detector (HD) gain is constant.

Using $\hat X$ and $\hat Y$, one can construct the joint quadratures
$\hat{X}_-=(\hat{X}_1-g\hat{X}_2)/\sqrt{2}$ and
$\hat{Y}_+=(\hat{Y}_1+g\hat{Y}_2)/\sqrt{2}$ needed to
calculate two measures of entanglement: the inseparability
\cite{Duan_inseperability,Simon_inseperability} and the extent to which our state satisfies the Einstein-Podolsky-Rosen (EPR) paradox
\cite{Reid_EPR}. To be applicable the two measures require the state to be Gaussian \cite{Nature_slowEPR}. The coefficient $g$ is a parameter which is used to optimize the entanglement measurement.

The inseparability criterion states that a necessary and sufficient
condition for entanglement is that
the sum of the joint quadrature operator variances satisfies:
$(\Delta \hat{X}_-)^2+(\Delta \hat{Y}_+)^2<(1+g^2)$ for some $g$.
Adjusting $g$ amounts to a local Bogoliubov transformation of the
uncertainty ellipses in phase space \cite{Duan_inseperability}, and
can be done in practice by electronically attenuating the signal
from one of the homodyne detectors, as shown in
Fig.~\ref{fig:setup}. We tune $g$ to minimize the joint quadrature
variances. When the variances are normalized to the corresponding
SQLs, the criterion becomes $I\equiv(\Delta \hat{X}_-)_N^2+(\Delta
\hat{Y}_+)_N^2<2$.

The EPR criterion indicates the extent to which a measurement on one
system (an optical mode) can give information about
the state of the other system \cite{Reid_EPR}. The EPR parameter is
obtained by measuring the conditional variances:
$E_{ij}\equiv V_{X_i|X_j}\cdot V_{Y_i|Y_j}$, where $V_{X(Y)_i|X(Y)_j}$ is
the variance of a prediction on a quadrature of system $i$, having
performed a measurement on system $j$. The criterion states that two
systems are EPR-entangled when $E_{ij} < 1$. One can also construct
the inequality by making a measurement on system $i$ and predicting
the result for system $j$ instead: $E_{ji}=V_{X_j|X_i}\cdot
V_{Y_j|Y_i}$. In general $E_{ji}\neq E_{ij}$
\cite{Bowen_entanglementPRA_2004}. The $V_{X(Y)_i|X(Y)_j}$ are the
joint quadrature variances normalized to the shot noise of the
system being estimated: $V_{X_1|X_2}=\Delta
(\hat{X}_1-g\hat{X}_2)^2|_{g=g_{\mathrm{min}}}$, where
$g_{\mathrm{min}}$ minimizes that variance \cite{Reid_EPR}, ensuring
that the joint quadrature measurement is done in the correct basis
\cite{Bowen_entanglementPRA_2004}. The $g$ that minimizes $\Delta
(\hat{X}_1-g\hat{X}_2)^2$ is different from its value when $I$ is
optimized. In both cases, the $g$ values were found empirically by
adjusting a variable electronic attenuator after the conjugate HD
(Fig.~\ref{fig:setup}).

Figure \ref{fig:squeezing}, left, shows the reference squeezing
level when the amplifier is turned off, and no ND filter is in the
conjugate beam path (the losses on the vapor cell windows still
affect the observed squeezing). The minima of each curve represent
the amplitude difference, $X_-$, and phase sum, $Y_+$, joint quadrature noises respectively.
After the first cell, both joint quadratures are squeezed equally
and both exhibit the same amount of antisqueezing, indicating the
initial twin beam generation process is phase insensitive. The right traces in
Fig.~\ref{fig:squeezing} show the squeezing level when the gain of
the amplifier is set to 1.8 and the attenuator transmits 56\% of the
light. The noise of both squeezed and antisqueezed quadratures
remains at equal levels, confirming that the amplification process is
also phase insensitive.

\begin{figure}[h] 
\includegraphics[width=3.0in]{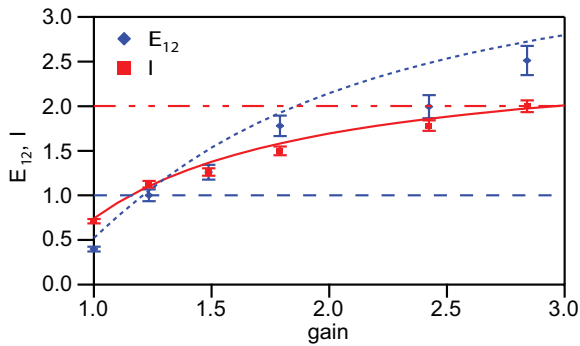}
\caption{(color online) $E_{12}$ and $I$ as a function of gain. The gain-loss product for each point was unity. The dash-dotted and dashed horizontal lines represent the upper bounds to the inseparability and EPR criteria, respectively. The solid curve shows the theoretical predictions for the inseparability, calculated from the total Hamiltonian of the amplifier-beamsplitter-detector system. The dotted curve shows the theoretical prediction for the EPR parameter, calculated following the method in \cite{Nature_slowEPR}. The error bars are propagated from a combined statistical and systematic uncertainty of 0.2~dB in the noise reduction measurements. The disagreement between $E_{12}$ and the theory is due to experimental difficulty in ascertaining the purity of the output state.} \label{fig:EPR_vs_atten}
\end{figure}

Figure \ref{fig:EPR_vs_atten} shows $E_{12}$ and $I$ as a function
of gain, where the ``12'' subscript represents a measurement on the
conjugate beam being used to predict the result of a measurement on
the probe beam.
One noteworthy aspect of Fig.~\ref{fig:EPR_vs_atten} is that the
state remains inseparable ($I<2$) for a gain of up to 2.8. It is
evident from the plot that a non-negligible degree of entanglement
still remains for a gain of 2 followed by 50\% attenuation. In other
words, after symmetrically cloning one mode from a two-mode squeezed state, the clones are entangled with the other unmodified mode from the
original state.

The EPR parameter reaches its limit of 1 with gain more quickly than
the inseparability. 
Nonetheless, up to a gain of 1.2, EPR correlations are
maintained. Given that the amplifier adds excess noise to the
conjugate, which unbalances the variances, the two EPR parameters
are not symmetric ($E_{12}\neq E_{21}$). Using the probe to infer a
property of the conjugate would be less successful, since only a
portion of the total conjugate noise is correlated with the probe
noise. The $E_{12}$ we have measured here represents the best case
scenario. The EPR parameter suffers much more from increasing gain than the inseparability because it is more sensitive to the degree to which the state is mixed. Inseparability describes how separable the density matrices of the two susbsystems are, immaterial of whether the total density matrix represents a pure state. As the gain increases and the ND filter transmission decreases, more and more excess noise  mixes with our original input, resulting in a mixed state. Because of this, we expect the EPR to degrade more quickly than the inseparability.

The excess noise added by the amplifier can be thought of as the
result of tracing over the unused ancilla, which is entangled with the amplified conjugate.  The
excess noise increases with gain, which leads to decreased quantum
correlations, as shown in Fig.~\ref{fig:EPR_vs_atten}. The quantum
entanglement is not totally lost, however. By measuring the ancillary beam we could extract more information about the entanglement between it and the cloned modes.

In this paper we have demonstrated the viability of 4WM in an atomic
vapor as a nearly quantum-noise-limited amplifier.
We have demonstrated the ability
to locally clone one mode from a two-mode squeezed state and
maintain entanglement. The resulting state would have entanglement distributed among three optical modes, and is reminiscent of the scheme for multipartite entanglement proposed by Ferraro, \textit{et al.}~\cite{Ferraro_mutipartiteJOSAB_2004}, while somewhat different from the multipartite entanglement schemes proposed by Pfister, \textit{et al.}~\cite{mutipartite_us_grouped}.
An interesting question is whether the method
described here is the best way to maintain entanglement while
cloning a single mode. It has been established that universal
amplifiers maintain the best fidelity for Gaussian states, but how
fidelity relates to entanglement when cloning only one mode is an
open question. Questions like these, along with the multiple cloning
ability of the device suggest many new avenues of exploration and
potential uses for quantum-noise-limited amplifiers based on 4WM. Further, the ability to amplify multiple spatial modes in parallel \cite{Science_images} can lead to the cloning of quantum images.

R.~C.~P.~was supported by a grant from the IC postdoctoral
research program.

\end{document}